\documentclass[reprint,pra]{revtex4-2}
\usepackage{graphicx}% Include figure files
\graphicspath{{Figures/}{}}

\usepackage{dcolumn}% Align table columns on decimal point
\usepackage{physics}
\usepackage{siunitx}
\AtBeginDocument{\RenewCommandCopy\qty\SI}
\usepackage{bm}% bold math
\usepackage[pdfstartview={FitH},pdfpagemode={UseNone}, breaklinks=true,colorlinks=true,bookmarks=false,allcolors=blue,bookmarksopen=true, pdfnewwindow=true]{hyperref}
\usepackage[all]{hypcap}

\usepackage[english]{babel}
\usepackage{amsmath}
\usepackage{hyperref}% add hypertext capabilities
\hypersetup{pdfstartview={FitH},pdfpagemode={UseNone}, colorlinks=true,bookmarks=false,allcolors=blue, breaklinks=true, bookmarksopen=true, pdfnewwindow=true}
\usepackage[all]{hypcap}

\usepackage[english]{babel}

\begin{document}

\title{Photon pair generation from lithium niobate metasurface \\ with tunable spatial entanglement}

\author{Jihua Zhang}
\email{jhzhanghust@gmail.com}
\author{Jinyong Ma, Dragomir N. Neshev}

\author{Andrey A. Sukhorukov}
\email{andrey.sukhorukov@anu.edu.au}

\affiliation{Centre of Excellence for Transformative Meta-Optical Systems (TMOS), Department of Electronic Materials Engineering (EME), Research School of Physics, The Australian National University, Canberra, ACT 2601, Australia}
%\affil{}
%\affil{Centre of Excellence for Transformative Meta-Optical Systems (TMOS), Australia}

% \corresp{\corsym{*}\emailLink{jhzhanghust@gmail.com}}
% \corresp{\corsym{**}\emailLink{andrey.sukhorukov@anu.edu.au}}

% \received{Month X, 2023}
% \accepted{Month X, XXXX}
% \posted{Month X, XXXX}

\begin{abstract}
Two-photon state with spatial entanglement is an essential resource for testing fundamental laws of quantum mechanics and various quantum applications. Its creation typically relies on spontaneous parametric down-conversion in bulky nonlinear crystals where the tunability of spatial entanglement is limited. Here, we predict that ultrathin nonlinear lithium niobate metasurfaces can generate and diversely tune spatially entangled photon pairs. The spatial properties of photons including the emission pattern, rate, and degree of spatial entanglement are analysed theoretically with the coupled mode theory and Schmidt decomposition method. We show that by leveraging the strong angular dispersion of the metasurface, the degree of spatial entanglement quantified by the Schmidt number can be decreased or increased by changing the pump laser wavelength and a Gaussian beam size. This flexibility can facilitate diverse quantum applications of entangled photon states generated from nonlinear metasurfaces.
%creating quantum photon states with tunable entanglement by using static metasurfaces.

% \OCIScodes{Spontaneous parametric
% down-conversion, metasurface, spatial entanglement.}

% \doi{10.3788/COLxxxxxx.xxxxxx}
\end{abstract}

\maketitle

\section{Introduction} \label{sec:intro}

\noindent Entanglement is an important feature of quantum mechanics that underpins various applications of quantum technologies \cite{Horodecki:2009-865:RMP}. In particular, photon pairs that are entangled in the high-dimensional spatial space represent an essential resource in a broad range of quantum applications including imaging \cite{Shih:2007-1016:ISQE, Moreau:2019-367:NRP}, communications \cite{Gisin:2007-165:NPHOT}, and computations \cite{Bennett:2000-247:NAT}. The performance of such
%these quantum 
applications is related to the degree of entanglement.
%of the photon pairs. 
The most common way to generate spatially entangled photon pairs is based on the spontaneous parametric down-conversion (SPDC) in quadratic nonlinear crystals, where a pump photon spontaneously splits into two lower-energy photons in two different directions \cite{Couteau:2018-291:CTMP}. The properties of generated states, including the emission pattern and spatial entanglement, 
%of the photon pairs from nonlinear crystals 
have been extensively investigated \cite{Law:2004-127903:PRL, Walborn:2010-87:PRP, Miatto:2012-263:EPD}. For nonlinear crystals with a typical thickness on the scale of millimeters to centimeters, the stringent phase matching condition limits the emission directions of the photon pairs to a certain predefined angle range, making it difficult to flexibly tune the spatial pattern and entanglement of the photon pairs while maintaining the generation efficiency. Although control of spatial correlations of the photon pairs was recently reported by engineering the pump beam profile \cite{Boucher:2021-4200:OL}, 
%the tuning range was relatively small and an increased pump power was needed to \comment{maintain the generation rate}. 
the specific tuning range of the spatial entanglement was unknown. It was shown that strong multi-mode entanglement can be achieved in thin nonlinear films~\cite{Okoth:2020-11801:PRA}, however the generation efficiency was much weaker compared to conventional schemes. Thereby, it remained a challenge on how to efficiently generate \textbf{tunable} spatially entangled photon pair.

Recently, it was shown experimentally that metasurfaces in the form of a nanostructured layers supporting optical resonances \cite{Chen:2020-604:NRM, Solntsev:2021-327:NPHOT, Li:2020-100584:PRSS, Chen:2021-823016:ACOS, Liu:2021-200092:OEA, Zhu:2022-327006:ACOS} can boost the photon pair generation \cite{Marino:2019-1416:OPT, Santiago-Cruz:2021-4423:NANL, Zhang:2022-eabq4240:SCA, Santiago_Cruz_2022}. 
%Due to the subwavelength thickness, the phase matching condition is relaxed. 
Furthermore, it was demonstrated that a lithium niobate metasurface featuring nonlocal guided-mode resonances can generate spatially entangled photon pairs in a broad angle range 
%with similar rate 
\cite{Zhang:2022-eabq4240:SCA}. 
%However, a systematic study on the structure of entanglement 
%spatial properties of the photon pairs generated from metasurfaces is still missing. 
In this work, we provide a theoretical description of the two-photon state generated from metasurfaces and a quantitative study on its degree of spatial entanglement, based on the Schmidt decomposition approach. 
By taking advantage of the strong angular dispersion of the nonlocal metasurface, we predict that the emission pattern of the photon pairs can be varied by simply changing the wavelength or beam size of the pump laser.
%without rate drop. 
%This property reveals a practical way to continuously vary the degree of spatial entanglement. The Schmidt decomposition of the emission pattern indicates that the Schmidt number can be tuned in a large range. 
This represents a simple and effective way to generate photon states with tunable spatial entanglement from ultra-thin metasurfaces, which may find applications in advanced quantum imaging and communications.

\section{SPDC from lithium niobate nonlocal metasurface} \label{sec:SPDC}

\begin{figure}
\centering	\includegraphics[width=0.96\columnwidth]{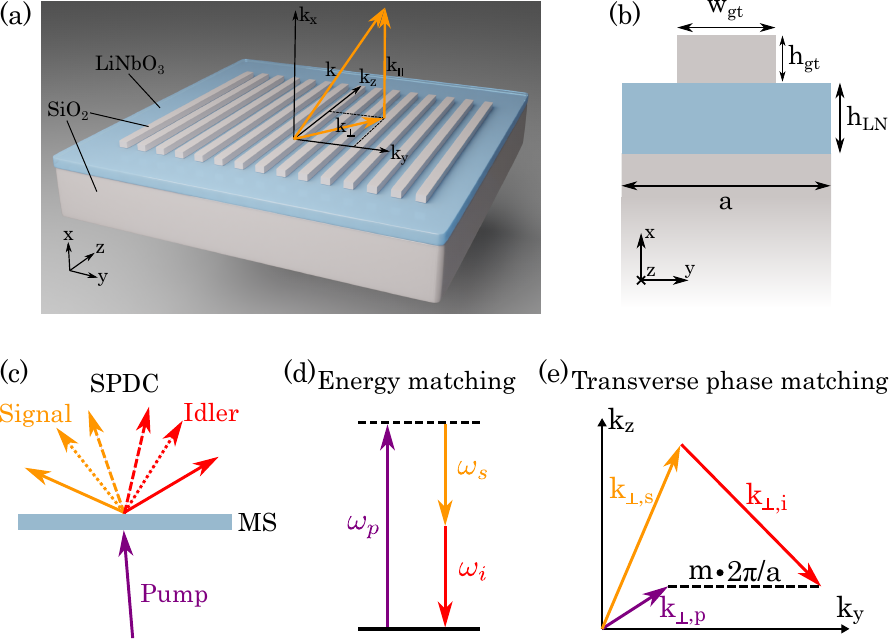}
\caption{(a) Schematic of the proposed metasurface consisting of a SiO$_2$ grating on top of a LiNbO$_3$ thin film on a SiO$_2$ substrate. Yellow arrow $\textbf{k}$ defines the wave vector of the emitted photons. %emission direction of the photons. 
$\textbf{k}_\bot$ and $\textbf{k}_\parallel$ are the transverse and longitudinal components of the wave vector, respectively. (b) Cross section of one unit cell of the metasurface. (c) Wide-angle emission of photons from the subwavelength-thick metasurface, satisfying (d) energy matching and (e) transverse phase matching.}\label{fig1}
\end{figure}

We consider the metasurface design following the experimental platform described in Ref.~\cite{Zhang:2022-eabq4240:SCA}. As shown in Figs.~\ref{fig1}(a,b), it is based on an \textit{x}-cut lithium niobate (LiNbO$_3$) thin film (thickness $h_{LN}=304\,$nm) covered by a silicon dioxide (SiO$_2$) grating with a period $a=890\,$nm, width $w_{gt}=550\,$nm, and thickness $h_{gt}=200\,$nm. The optical axis of the LiNbO$_3$ and the grating are along the in-plane $z$ direction. In the SPDC process, a pump photon splits into two photons called signal and idler photons. This metasurface supports nonlocal guided mode resonance at the signal and idler wavelengths around 1570~nm, which facilitates an increased density of states 
%for the signal and idler photons 
and boosts the SPDC process. 

Importantly, due to the subwavelength thickness of the metasurface, the longitudinal phase matching is relaxed, allowing SPDC in a broad anglular range, as shown in Fig.~\ref{fig1}(c). The frequencies and wave vectors of the photons satisfy the following energy and transverse phase matching conditions, as illustrated in Figs.~\ref{fig1}(d,e):
\begin{equation}
    \omega_s+\omega_i=\omega_p \, ,
    \label{eq:em}
\end{equation}
\begin{equation}
    \textbf{k}_{\bot,s}+\textbf{k}_{\bot,i}+m\cdot2\pi/a \hat{\bf{y}}= \textbf{k}_{\bot,p} \, ,
    \label{eq:pm}
\end{equation}
where $\omega_{p}$, $\omega_{s}$ and $\omega_{i}$ are the angular frequencies of the pump, signal and idler photons, $\textbf{k}_{\bot,p,}$, $\textbf{k}_{\bot,s}$ and $\textbf{k}_{\bot,i}$ are their transverse wave vectors in the $y-z$ plane, and $m$ is an integer. For a specific pump, matching these two conditions means that if the frequency or angle of one photon is known, the information for another photon is determined automatically. This essentially leads to the frequency and spatial correlations of the photon pairs. In the following, we consider the case of $m=0$ corresponding to the experimental conditions in Ref.~\cite{Zhang:2022-eabq4240:SCA}.  The resulting two-photon state has a wavefunction
\begin{equation}
\begin{aligned}
    |\Psi\rangle = \int d\omega_s d\omega_i d\textbf{k}_{\bot,s} d\textbf{k}_{\bot,i} S(\omega_s+\omega_i, \textbf{k}_{\bot,s} + \textbf{k}_{\bot,i}) \\
    \Xi(\omega_s,\textbf{k}_{\bot,s};\omega_i,\textbf{k}_{\bot,i})|\omega_s,\textbf{k}_{\bot,s}\rangle |\omega_i,\textbf{k}_{\bot,i}\rangle \, ,
\end{aligned}
    \label{eq:wf}
\end{equation}
where $S$ is the normalized frequency-angular spectrum of the pump light. $\Xi(\omega_s,-\textbf{k}_{\bot,s};\omega_i,-\textbf{k}_{\bot,i})$ is the SPDC efficiency function.

The wavefunction is controllable by the spectrum of the pump, similar to the case in conventional nonlinear crystals \cite{Boucher:2021-4200:OL}. Note that this control is weighted by the SPDC efficiency function. The difference in metasurfaces is that the SPDC is free from longitudinal phase matching and its efficiency is fully determined by the frequency and angular dispersion of the optical resonances. This feature enables control of the spatial properties of the photon pairs by engineering optical resonances supported by the metasurface.
%in metasurfaces with nonlinear layers of sub-wavelength thickness. 
For example in the proposed metasurface with angular dependent nonlocal resonances, the photon-pair emission pattern is very sensitive to the frequency and angular spectra of the pump. In contrast, the SPDC efficiency in bulky nonlinear crystals is limited by the longitudinal phase matching, being proportional to $\text{sinc}(|\delta \textbf{k}_\parallel| L/2)$, where $L$ is the thickness of the crystal and $\delta \textbf{k}_\parallel = \textbf{k}_{\parallel,p}-(\textbf{k}_{\parallel,s}+\textbf{k}_{\parallel,i})$ is the longitudinal phase mismatch \cite{Monken:1998-3123:PRA, Unternahrer:2018-1150:OPT}.
%\comment{Because of this reason,} the emission pattern of the photon pairs \comment{from the bulky crystals is not tunable, typically determined by the phase matching techniques and the thickness of the crystals.} %typically to a ringed pattern \comment{depends on the phase matching techniques}, which depends on the the thickness of the crystal. 
%In contrast, the SPDC efficiency in metasurfaces is free from longitudinal phase matching and it is rather determined by the frequency and angular dispersion of the optical resonances 
%\comment{the photon emission manifests strong angular dispersion induced by the nonlocal resonance}. This feature \comment{enables} control of the spatial properties of the photon pairs by engineering the optical \comment{properties of the pump laser.} %resonances 
%in metasurfaces with nonlinear layers of sub-wavelength thickness. 
%without changing the thickness.

\section{Quantum-classical correspondence}
\begin{figure}
\centering
	\includegraphics[width=0.96\columnwidth]{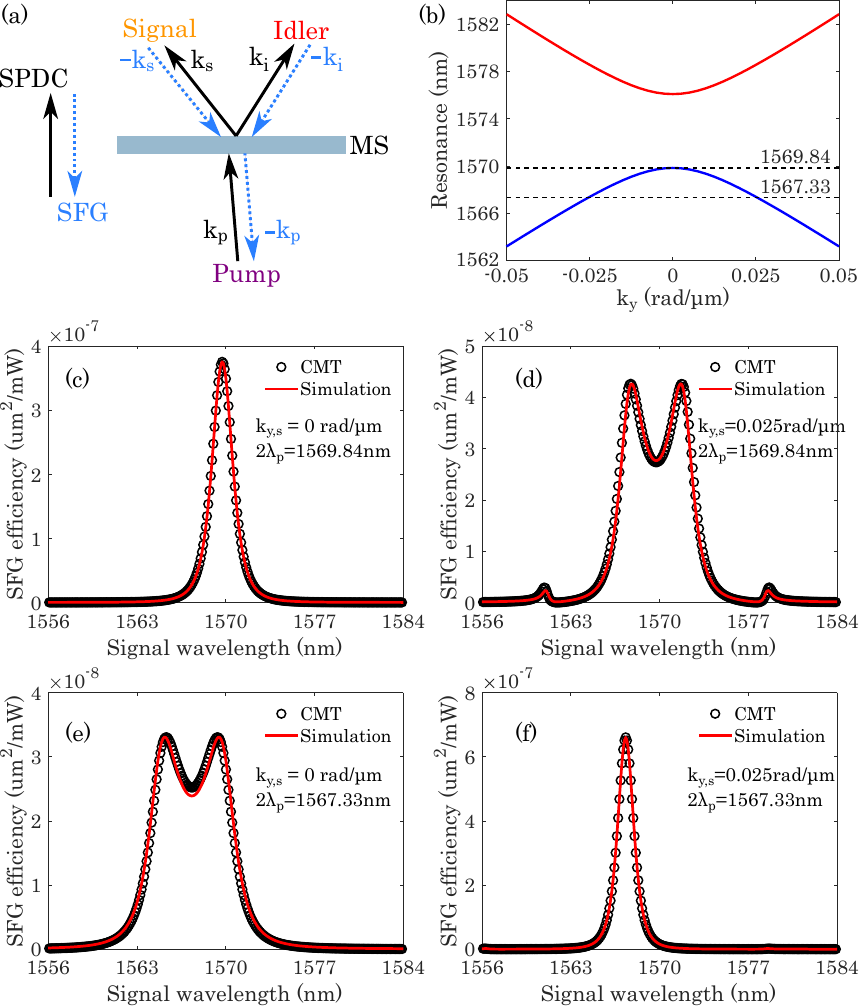}
\caption{(a) Quantum-classical correspondence between the SPDC and SFG. (b) CMT predicted resonance wavelengths of the metasurface as a function of the transverse wave vector $k_y$ at $k_z=0$. (c-f) SFG efficiency as a function of the signal wavelength calculated by CMT (black circles) and COMSOL simulation (red lines) for different input transverse wave vectors of the signal ($k_{y,s}=0$, $0.025 \;rad/\mu$m) and wavelengths of the pump. Double of the pump wavelength is (c,d)~1569.84~nm and (e,f)~1567.33~nm, which are marked by the black dashed lines in (b). 
%\comment{add a dashed line in (c)-(f) to indicate the resonance}
}
\label{fig2}
\end{figure}

The theoretical modelling of quantum photon-pair generation throigh SPDC in quadratically nonlinear metasurfaces can be performed through the general Green's function formalism~\cite{Poddubny:2016-123901:PRL}.
%As SPDC is a quantum process, it is hard to directly model it in metasurfaces. 
A mathematically equivalent approach, which can be more convenient for numerical modelling, is based on
%To solve this issue, we apply 
the quantum-classical correspondence [Fig.~\ref{fig2}(a)], where the SPDC efficiency is derived from its classical reverse process called sum-frequency generation (SFG) \cite{Helt:2015-1460:OL, Lenzini:2018-17143:LSA, Marino:2019-1416:OPT, Kaneda:2020-38993:OE, Parry:2021-55001:ADP, Zhang:2022-eabq4240:SCA}:
\begin{equation}
    \Xi(\omega_s,\textbf{k}_{\bot,s};\omega_i,\textbf{k}_{\bot,i}) = \sqrt{\frac{\omega_s\omega_i}{\omega_p^2(2\pi)^3}} \xi_{SFG}(\omega_s,-\textbf{k}_{\bot,s};\omega_i,-\textbf{k}_{\bot,i}) \, .
    \label{eq:qcc}
\end{equation}
Here $\xi_{SFG}(\omega_s,-\textbf{k}_{\bot,s};\omega_i,-\textbf{k}_{\bot,i})=A_p/(A_s A_i)$ is the SFG efficiency in the reverse direction with $A_p$ being the output SFG and $A_{s}$ ($A_{i}$) being the input signal (idler) complex amplitudes of unit-intensity plane-waves. 
%$I_{s,i}$ are set to match $I_s=I_i\omega_s/\omega_i$ so that the numbers of input signal and idler photons are the same. 
In this work, we are interested in the angular or spatial properties of the photons and consider a continuous-wave pump at normal incidence. Here we focus on the case of degenerate SPDC ($\omega_s=\omega_i=0.5\omega_p$) since the frequency spectrum of photons was found to be narrow for a normally incident pump~\cite{Zhang:2022-eabq4240:SCA}. This is because the degenerate SPDC efficiency is much stronger than in the non-degenerate case, as we show in the following Sec.~\ref{sec:CMT}. The corresponding spatial wavefunction becomes
\begin{equation}
\begin{aligned}
    |\Psi\rangle = \int d\textbf{k}_{\bot,s} & d\textbf{k}_{\bot,i} \frac{S(\textbf{k}_{\bot,s} + \textbf{k}_{\bot,i})}{4\pi} \\
    & \xi_{SFG}(-\textbf{k}_{\bot,s};-\textbf{k}_{\bot,i}) 
    |\textbf{k}_{\bot,s}\rangle |\textbf{k}_{\bot,i}\rangle
    \label{eq:wf1}
    \end{aligned}
\end{equation}
After calculating this two-photon wavefunction, we can reveal the emission brightness and pattern of the SPDC, and the spatial entanglement of the photon pairs. Apparently, they are controllable through the angular spectrum of the pump and the angular-dependent SFG efficiency determined by the optical resonance of the metasurface. The complexity of Eq.~(\ref{eq:wf1}) comes from the fact that for a specific pump, the photon pairs from the metasurface can emit to all directions satisfying the transverse phase matching conditions in Eq.~(\ref{eq:pm}).
As a result, one needs to simulate the SFG process in all directions to simulate the full SPDC process, which is a computationally demanding 
%time-consuming 
task. 

\section{Coupled mode theory for SPDC modelling} \label{sec:CMT}

To efficiently model the SPDC process in the lithium niobate metasurface, we have proposed a coupled mode theory (CMT) that can accurately calculate the frequency and angular dispersion of the guided mode resonances in the metasurface (see detailed theory in the Supplementary of Ref.~\cite{Zhang:2022-eabq4240:SCA}). Figure~\ref{fig2}(b) shows the CMT predicted resonance wavelengths of the metasurface at different transverse wave vectors along the $y$ direction. Based on the CMT, we can also calculate the SFG efficiencies for different $k_{y,s}$ and pump wavelength $\lambda_p$. A normally incident plane-wave pump is considered in the calculations. As shown in Figs.~\ref{fig2}(c-f), the CMT results show a good agreement with the full-wave simulations conducted in COMSOL Multiphysics, confirming its high accuracy in modeling SFG and the coresponding SPDC process based on Eq.~(\ref{eq:wf1}). Importantly, the COMSOL simulation takes around one hour to calculate 500 points of the signal wavelength, while the CMT takes 14 milliseconds only. Note that the SFG efficiencies in Figs.~\ref{fig2}(c,f) are about an order of magnitude higher than the ones in Figs.~\ref{fig2}(d,e). This is because in Figs.~\ref{fig2}(c,f) the degenerate signal and idler photons are both at resonance wavelength, while only either a signal or idler photon is resonant for the non-degenerate case in Figs.~\ref{fig2}(d,e). Due to the quantum-classical correspondence, we expect that the degenerate SPDC efficiency is the strongest in the specific direction  determined by the pump wavelength. Therefore, only the frequency-degenerate SPDC is considered in the following analysis of the spatial entanglement of the photon pairs. 

%This is because in these two configurations, the resonance wavelength of the metasurface at the related wave vectors equal to two times of the applied pump wavelength. As a result, the degenerate signal and idler light, which are input from the opposite transverse wave vectors, resonate at the same time. This double resonance is manifested by a single and enhanced SFG peak at the resonance wavelength. Due to the correspondence, we expect that the SPDC efficiency is also the strongest in the related directions when pumping at specific wavelengths.

\section{Quantification of spatial entanglement} \label{sec:entang}

\begin{figure}
\centering
	\includegraphics[width=0.96\columnwidth]{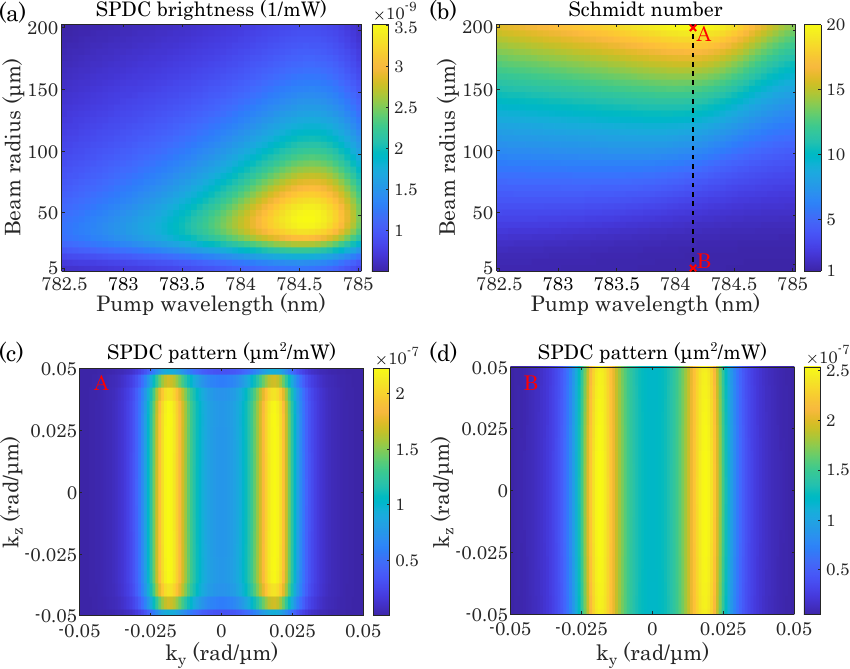}
\caption{(a) SPDC brightness at the degenerate wavelength and (b) Schmidt number of the emitted photons as a function of the pump laser wavelength and Gaussian beam radius. Point $A$ in (b) marks the peak Schmidt number at pump wavelength of 784.15 nm and beam radius of 200 $\mu$m. Point $B$ corresponds to the same pump wavelength with a beam radius of 5 $\mu$m. (c,d) SPDC emission patterns corresponding to the points $A$ and $B$ in (b), as indicated by labels.}
\label{fig3}
\end{figure}
%

%For a pump with a specific wavelength and angular spectrum, 
We now calculate SFG at the degenerate wavelength for different wave vectors and obtain the two-photon wavefunction of the photons based on Eq.~(\ref{eq:wf1}). Specifically, we consider a Gaussian pump beam with an angular spectrum $S(\textbf{k}_\bot)=\exp(-\sigma_p^2|\textbf{k}_\bot|^2/4)$, where $\sigma_p$ is the beam waist radius. By integrating over the wave vector, we obtain the spectral brightness of the SPDC, as shown in Fig.~\ref{fig3}(a) for different pump wavelengths and beam radii. In the calculation, an upper bound of 0.05~$rad/\mu$m has been applied for $k_y$ and $k_z$, which corresponds to a collection angle of 0.7~degrees in previous experimental setup~\cite{Zhang:2022-eabq4240:SCA}. One can see that for each beam size there is a peak for the SPDC at 1569.1~nm, which is slightly blue shifted from the resonance at normal incidence. %and consist with the results in \cite{Zhang:2022-eabq4240:SCA}. 
We notice that the highest brightness occurs at $\sigma_p=45 \;\mu$m. This is because the generation rate of photon pairs depends on the incident angle of the plane-wave pump 
%The weight of the wave vector components decomposed from the pump beam with a radius of $\sigma_p=45 \;\mu$m is found to be peaked at $k_{y,p}=0.015 \;rad/\mu$m where the generation rate is maximized (See Fig.~S3d in the Supplementary of Ref.~\cite{Zhang:2022-eabq4240:SCA}). 
and the maximum happens near $k_{y,p}=0.015 \;rad/\mu$m, as shown by Fig.~S3d in the Supplementary of Ref.~\cite{Zhang:2022-eabq4240:SCA}. The pump beam with a radius of $\sigma_p=45 \;\mu$m has the largest weight at the wave vector components near this value. 

\begin{figure}
\centering
	\includegraphics[width=0.96\columnwidth]{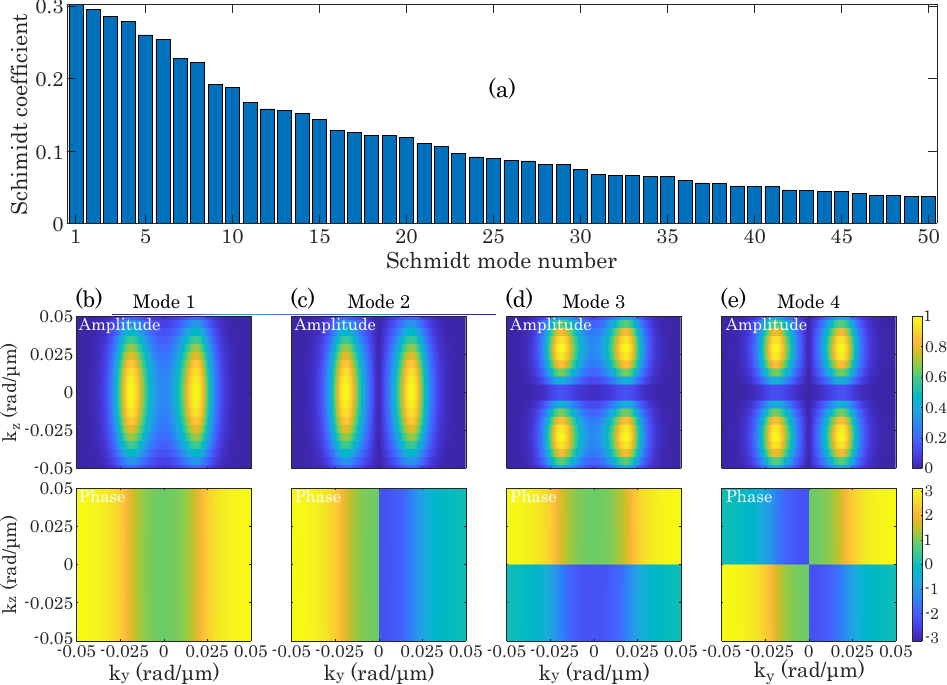}
\caption{(a) Schmidt coefficients of the first 50 Schmidt modes corresponding to the point $A$ in Fig.~\ref{fig3}(b). (b-e) The normalized amplitude and phase distributions of the first four Schmidt modes.}\label{fig4}
\end{figure}

We perform a Schmidt decomposition of the two-photon wavefunction to quantify the spatial entanglement. The related Schmidt numbers are shown in Fig.~\ref{fig3}(b).  Obviously, the Schmidt number is strongly dependent on the wavelength and beam size of the pump and can be tuned in a large range from 1.1 (weakly-entangled) to 20 (strongly entangled). A larger beam radius leads to a larger Schmidt number. This is reasonable since a plane-wave pump with an infinite beam radius has a single transverse wave vector, resulting in photon pairs with fully deterministic relation between the signal and idler wavevectors according to the transverse phase matching, and each of a continuum number of such pairs represents a Schmidt mode. For the beam radius of 200 $\mu$m, the maximum Schmidt number is found to be 20 when two times of the pump wavelength is 1568.3~nm, as marked by point $A$ in Fig.~\ref{fig3}(b). At the same pump wavelength, the Schmidt number is only 1.1 for a pump radius of 5 $\mu$m, which is point $B$ in Fig.~\ref{fig3}(b). The related SPDC emission patterns for these two points are shown in Figs.~\ref{fig3}(c) and~(d), respectively. 
%For point $A$, the emission has two separate areas near $k_y=\pm 0.017 rad/\mu$m and shows a weak dependence on $k_z$. While for point $B$, the emission pattern becomes larger along the $k_y$ direction and are no longer constant along $k_z$ direction. This is due to the appearance of large-magnitude wave number components. 

\begin{figure}
\centering
	\includegraphics[width=0.96\columnwidth]{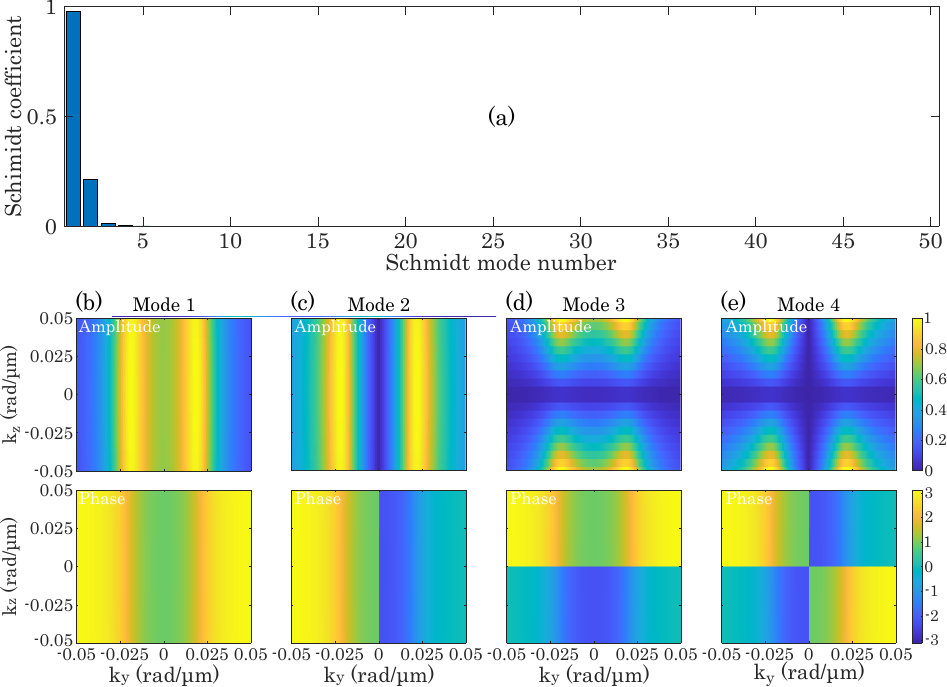}
\caption{(a) Schmidt coefficients of the first 50 Schmidt modes corresponding to the point $B$ in Fig.~\ref{fig3}(b). (b-e) The normalized amplitude and phase distributions of the first four Schmidt modes.}\label{fig5}
\end{figure}

To gain more insight of the spatial entanglement in Fig.~\ref{fig3}(b), we focus on the Schmidt decomposition results at points $A$ ($2\lambda_p=1568.3 \text{ nm}, \sigma_p=200 \;\mu$m) and $B$ ($2\lambda_p=1568.3\text{ nm}, \sigma_p=5\;\mu$m). Figure \ref{fig4} depicts the Schmidt coefficients of the lowest 50 Schmidt modes and the mode distributions of the lowest four modes for point $A$. As can be seen, all the 50 Schmidt coefficients are nonzero and the first four modes have similar magnitudes, indicating a high degree of spatial entanglement. The results for point $B$ in Fig.~\ref{fig3}(b) are shown in Fig.~\ref{fig5}. The Schmidt coefficients show a fast drop for higher-order modes and are close to zero for the modes with mode number over 3, confirming a week entanglement at point $B$. %This explains the small Schmidt number of point $B$.

% \vspace{-0.5 cm}
\section{Conclusions} \label{sec:concl}
In conclusion, we have theoretically investigated the generation of spatially entangled photon pairs via spontaneous parametric down-conversion a lithium niobate metasurface. By quantifying the spatial entanglement with the Schmidt number, we have shown that the degree of spatial entanglement can be diversely tuned by the pump laser wavelength and beam size. The capability to realize arbitrary spatial entanglement can find applications in quantum imaging, which resolution is related to the spatial entanglement of the photon source. In the future, pump beams with tailored spatial profiles and metasurfaces with different optical resonances can be further explored to tune both the emission pattern and spatial entanglement.
%
% \vspace{-0.5 cm}
\section*{Acknowledgement}
This work was supported by the Australian Research Council (DP190101559, CE200100010).
\bibliography{db_COL-invited-JZ}
\end{document}